\documentclass[runningheads]{llncs}
\usepackage[T1]{fontenc}
\usepackage{graphicx}
\usepackage{booktabs} % For \toprule, \midrule, \bottomrule
\usepackage{amsmath}

\begin{document}
\title{Extending the Frontiers of QNLP Beyond English: Grammar-Sensitive Pipeline for Hindi Sentiment Classification Using Compositional Quantum Models}
\titlerunning{QNLP Hindi Sentimental Analysis}
\author{
Gautami Sanjay Naik\inst{1} \and
Rishi Koushik Reddy Thippireddy\inst{1} \and
Naman Srivastava\inst{2} \and
Parishri Shah\inst{1} \and
Ravi Raj\inst{1} \and
Sunil Saumya\inst{1} \and
Aswath Babu H\inst{1}
}
\authorrunning{Naik et al.}
\institute{
Indian Institute of Information Technology, Dharwad, Karnataka, India\\
\email{\{gautaminaik2000, rishikoushik18, parishrishah24, atomraviraj199\}@gmail.com,\{sunil.saumya, aswath\}@iiitdwd.ac.in} \and
Indian Institute of Science, Bengaluru, Karnataka, India\\
\email{srinaman2@gmail.com}
}

\maketitle              % typeset the header of the contribution
\begin{abstract}
Advancements in Natural Language Processing (NLP), whether via classical or quantum platforms, have predominantly focused on English due to its global usage and the abundance of linguistic resources. Although English remains the most studied language in computational linguistics, Hindi, the third most spoken language worldwide after Mandarin, has received comparatively limited attention. Spoken primarily in India, Hindi differs significantly from English in its script, syntactic structure, and cultural-linguistic context. Linguistically, English belongs to the Germanic branch of the Indo-European family, whereas Hindi, derived from Sanskrit, is part of the Indo-Aryan branch. It uses the Devanagari script, exhibits rich morphological inflection, and follows a subject–object–verb (SOV) order, in contrast to English’s relatively simpler morphology and subject–verb–object (SVO) structure. These typological differences make Hindi a compelling candidate for exploring the generalization capacity of quantum-enhanced NLP models. Motivated by Hindi’s syntactic and morphological complexity and its underrepresentation in both classical and quantum NLP research, we propose a grammar-aware Quantum NLP (QNLP) pipeline for Hindi sentiment classification. A key focus is on handling sentential negation, a linguistic feature inadequately addressed in current models. Most QNLP implementations are limited to English and lack grammar-sensitive methods for morphologically rich and syntactically flexible languages. We use a manually annotated dataset of Hindi sentences labeled as positive, negative, or neutral, and encode them using pregroup grammar types. Sentences are processed through Lambeq, generating quantum circuits using a novel, negation-aware compositional grammar. We train Hybrid Quantum Neural Networks (HQNNs) for both binary and ternary sentiment classification. Our results demonstrate successful classification performance and showcase QNLP’s potential for typologically diverse languages.

\end{abstract}
\section{Introduction}
Quantum Natural Language Processing (QNLP) represents an emerging paradigm that combines quantum computing with natural language processing (NLP) to encode linguistic meaning into quantum states for language task execution \cite{Meichanetzidis_2021}.  The Distributional Compositional Categorical (DisCoCat) model of meaning developed by Coecke {\em et al.} (2010) \cite{coecke2010mathematical} serves as the foundation for QNLP to create a mathematical system that unites grammatical structures with semantic meaning. The DisCoCat framework uses Lambek pregroups as grammar types to determine how word representations (in terms of quantum states or tensors) merge and evolve as sentence meanings~\cite{lambek2006pregroups}. The compositional nature of this approach aligns perfectly with quantum principles as tensor mathematics for meaning directly corresponds to quantum mechanics formalism. Theoretical research at an early stage indicated that QNLP could achieve algorithmic benefits through quadratic speed-up for particular operations by representing the linguistic structure as quantum states.

The field of QNLP has transitioned from theoretical development to practical implementation through recent advancements.  The availability of noisy intermediate-scale quantum (NISQ) hardware allows researchers to conduct initial experiments that run NLP models as quantum circuits \cite{kartsaklis2021lambeqefficienthighlevelpython}. Lorenz {\em et al.} (2021) \cite{10.1613/jair.1.14329} achieved the first execution of small QNLP models on operational quantum computers, wherein the authors used an IBM quantum device to run their QNLP pipeline, for basic sentence meaning binary classification while demonstrating positive results despite hardware constraints. So far research on using QNLP to analyze sentiment in English language content has been perceived~\cite{ruskanda2023simple,9776836}. The two-class sentiment classification system developed by Lorenz {\em et al.} \cite{10.1613/jair.1.14329}  was expanded by Martínez {\em et al.} (2022) \cite{martinez2022multiclass} to perform emotion classification on a bigger dataset with four emotional categories. The DisCoCat QNLP framework demonstrated its ability to handle real-world applications by achieving high accuracy on small test datasets.

 Most previous QNLP implementations have focused on English-language datasets, but multilingual adaptation has been minimally researched. Previous studies were insufficient in addressing the computational complexities encountered after selection due to the necessity of precise measurement outcomes for quantum circuits in order to achieve reliable results.
 
 This paper presents an application of the QNLP method for Hindi sentiment classification. The Hindi language has a head-final structure with a Subject-Object-Verb (SOV) order, complex morphology, and flexible word order, which makes it difficult for QNLP pipelines that were originally designed for English  (SVO order). To address this, we manually developed a Hindi pregroup grammar resource, assigning types to words such that sentences are reduced grammatically. While there are preliminary efforts to formalize Hindi grammar using pregroups \cite{akshar1995natural}, there was no off-the-shelf parser available for Hindi. We used the open-source Lambeq toolkit \cite{Meichanetzidis_2021} to convert the manually typed Hindi sentences into quantum tensor networks and parameterized quantum circuits.
 
 The syntactic structure of each Hindi sentence is converted into string diagrams, which are then translated into parameterized quantum circuits using the Lambeq toolkit. The Hybrid Quantum Neural Network (HQNN) architecture is used to classify the sentiment of each sentence into two categories: positive and neutral. The circuit construction and training procedure are explained in detail in the Methodology section.

\section{Background}
\subsection{Foundations of QNLP}QNLP receives its foundation from the categorical compositional model of meaning developed by Coecke {\em et al.} (2010) \cite{coecke2010mathematical}, which demonstrated how grammatical composition can be translated into linear algebra and tensor  products. The semantic vector space contains each word of a sentence as a vector that functions as a quantum state, while pregroup or categorial grammar determines the vector combination rules for sentence-level vector creation. The connection between grammar and quantum mechanics emerges because grammatical reductions can be expressed through string diagrams that equal quantum circuits in the mathematics of the tensor network \cite{kartsaklis2021lambeqefficienthighlevelpython}. The DisCoCat framework allows compositional meaning computation as it uses tensor contractions (via quantum  circuits) to calculate sentence meanings based on syntactical word meaning interactions. The work of Coecke {\em et al.} introduced a new approach to handle natural language semantics through quantum formalisms, which showed that quantum computers would execute these operations naturally. The concept received further development through the Foundations for Near-Term QNLP by Coecke and coworkers (2020)~\cite{Meichanetzidis_2021}, who adapted DisCoCat to NISQ device-friendly quantum circuits while discussing the integration of quantum machine learning with compositional linguistics.

\subsection{QNLP Tooling (DisCoPy and Lambeq)} Research has been undertaken to come up with software frameworks to automate the process of converting sentences into quantum computations for practical QNLP model implementation. DisCoPy  (Distributional Compositional Python) is a library for string diagrams and monoidal categories that offers a formal method to work with the graphical structures that underpin QNLP \cite{martinez2022multiclass}. It enables the representation of grammatical reductions through diagrammatic operations and enables the connection to quantum circuit libraries. Kartsaklis {\em et al.} (2021) \cite{kartsaklis2021lambeqefficienthighlevelpython} developed Lambeq as a high-level QNLP toolkit. The QNLP pipeline is fully supported by Lambeq, which starts with sentence parsing through A Combinatory Categorial Grammar (CCG) parser via pregroup grammar, which further leads to string diagram generation and parameterized quantum circuit creation. The Lambeq generates quantum circuit ansatze from the syntax trees of sentences through qubit assignment and rotation gate allocation for words and entangling operations for grammatical connections. The toolkit provides “QNLP-in-practice” experiments through its ability to hide technical details while following the same processing sequence as Meichanetzidis {\em et al.} and Lorenz {\em et al.} in their first QNLP hardware experiments. Our research work uses Lambeq together with DisCoPy to process Hindi sentences by inputting manually parsed Hindi sentences into Lambeq, which produces simplified string diagrams and quantum circuits that we modify for our HQNN model.

\subsection{Prior QNLP experiments on English}
The first demonstrations of QNLP focused on English as parsing tools are readily available for this language. Lorenz {\em et al.} (2021) performed binary sentiment classification on real quantum hardware using Lambeq-generated circuits, but they heavily depended on post-selection and templated English sentences \cite{10.1613/jair.1.14329}. Ganguly {\em et al.} (2023) employed Lambeq for binary sentiment classification on a small dataset [5], and Martinez and Leroy-Méline (2022) expanded QNLP to four emotion classes using synthetic English data \cite{martinez2022multiclass}. These early implementations primarily addressed English-based datasets without extending the framework to structurally diverse languages.

\subsection{Pregroup Grammar} The algebraic framework of pregroup grammar was first introduced by Lambek \cite{lambek2006pregroups}. The system includes atomic types and their adjoint (left/right) counterparts, which enable the recognition of grammatical sentences through the cancelation of matching adjoint types. The formalism has been used to represent the syntax of many languages, including English, French, Japanese, and Sanskrit, Arabic as a single uniform system~\cite{bargelli2001algebraic,casadio2009clitic,cardinal2002algebraic,casadio2014word}. Coecke {\em et al.} (2010) showed that  pregroups function as compositional building blocks for semantic models, including quantum natural language processing, through type alignment with tensor structures~\cite{debanth2023computational,10.1007/3-540-48975-4_1}.

With regards to Hindi, the analysis of the formal structure faces multiple challenges because of its typological features. Here the language structure features head-finality and follows an SOV pattern with postpositional case markers. The word order follows the Subject-Object-Verb, while determiners and case markers (postpositions) occur after the noun and verbs (including auxiliaries) appear at the end of the clause. The postpositional structure with the SOV word order makes it difficult to directly use the grammatical frameworks that were developed for English (SVO and prepositional). The word order of Hindi shows flexibility because constituents can be rearranged for emphasis or topicalization, but these rearrangements follow specific grammatical rules. These properties require a grammar system that can process meaningful word-order changes while maintaining proper syntactic relationships~\cite{debanth2023computational}.
 
Shrivastava and Debnath (2021) proposed a pregroup formalization for Hindi using Paninian karaka roles~\cite{10.1007/3-540-48975-4_1}. Each case-marked noun phrase is typed according to its syntactic function: subjects as $k_1$, objects as $k_2$, and so on. These combine with postpositions and verbal arguments to reduce to the sentence types \cite{palmer2009hindi}. Further taking inference from these references, we have come up with a Pregroup Grammar Types for Hindi-English Compositional Semantics as shown in the Table.~\ref{tab:pregroup-grammar}.

\begin{table}[h]
\centering
\resizebox{\columnwidth}{!}{%
\begin{tabular}{@{}c l l@{}}
\toprule
\textbf{Symbol} & \textbf{Category} & \textbf{Meaning} \\ \midrule
$s$ & Sentence & Complete sentence \\
$n$ & Noun & General noun type \\
$o$ & Object & Object of verb \\
$p$ & Predicate & Verb or verb phrase \\
$\pi_{11}, \pi_{12}, \pi_{13}$ & Pronoun (Singular) & First, Second, Third person pronouns \\
$\pi_{21}, \pi_{22}, \pi_{23}$ & Pronoun (Plural) & First, Second, Third person plural pronouns \\
$\tau_{1}, \tau_{2}, \tau_{3}$ & Tense Marker & Past, Present, Future \\
$\kappa_{1}$ to $\kappa_{8}$ & Karaka Marker & \textit{ne}, \textit{ko} (object), \textit{se}, \textit{ko} (dative), \textit{ke liye}, \textit{se} (source), \textit{me}, \textit{par} \\
$\rho_{1}, \rho_{2}$ & Sambandha Marker & Masculine, Feminine agreement \\
$\alpha_{1}, \alpha_{2}, \alpha_{3}$ & Aspect Marker & Perfect, Imperfect, Continuous \\
$n_{11}, n_{12}$ & Singular Nouns & Masculine, Feminine \\
$n_{21}, n_{22}$ & Plural Nouns & Masculine, Feminine \\
\bottomrule
\end{tabular}%
}
\caption{Pregroup Grammar Types for Hindi-English Compositional Semantics}
\label{tab:pregroup-grammar}
\end{table}

Hindi’s verb phrases are complex, often involving aspect markers (such as {\em raha, chuka}) and tense auxiliaries (such as {\em hai, tha}) as shown in the Table~\ref{tab:nahi-structures}. The pregroup system handles this via intermediate types like {\em aspect} and {\em tense} denoted as $\alpha$ and  $\tau$, respectively. A typical transitive verb like  khata (“eats”) would take subject and object roles as inputs and expect a tense marker to yield sentences `s'. For instance, take a Hindi sentence "main seb khata hu" ({\em I eat an apple}), the structure reduces to "s" through valid contractions~\cite{debanth2023computational}.
\begin{equation}
    \pi~ o ~ o^r@\pi^r@s@\tau^l ~ \tau. 
\end{equation}

\section{Dataset Collection}

\begin{table}[hbtp]
\centering
\resizebox{\textwidth}{!}{
\begin{tabular}{@{}lll@{}}
\toprule
\textbf{Sentence Structure} & \textbf{Example (Hindi $\rightarrow$ English)} & \textbf{Pregroup Type Sequence} \\ \midrule
Simple Affirmative (SOV) & Ram gaya. (``Ram went.'') & $n_{11}^{~} \quad n_{11}^r@s$ \\[0.5em]
Simple Negation (SOV) & Ram nahi gaya. (``Ram did not go.'') & $n_{11} \quad (n_{11}^r@s) @ (n_{11}^r@s)^l \quad n_{11}^r @ s$ \\[0.5em]
With Auxiliary (Affirmative) & Main seb khata hu. (``I eat an apple.'') & $\pi_{11} \quad o \quad o^r @ \pi_{11}^r @ s @ \tau_2^l \quad \tau_2$ \\[0.5em]
With Auxiliary (Negation) & Main seb nahi khata hu. (``I do not eat an apple.'') & $\pi_{11} \quad o \quad (o^r @ \pi_{11}^r @ s @ \tau_2^l) @ (o^r @ \pi_{11}^r @ s @ \tau_2^l)^l \quad o^r @ \pi_{11}^r @ s @ \tau_2^l \quad \tau_2$ \\[0.5em]
Progressive Negation (Pre-verb) & Main nahi kha raha hu. (``I am not eating.'') & $\pi_{11} \quad (o^r @ \pi_{11}^r @ p @ \alpha_3^l) @ (o^r @ \pi_{11}^r @ p @ \alpha_3^l)^l \quad o^r @ \pi_{11}^r @ p @ \alpha_3^l \quad \alpha \quad \tau^l \quad \tau$ \\[0.5em]
Progressive Negation (Pre-auxiliary) & Main kha nahi raha hu. (``I am not eating.'') & $\pi_{11} \quad o^r @ \pi_{11}^r @ p @ \alpha_3^l \quad (\alpha_3 \alpha_3^l)^l \quad \alpha_3 \quad \tau_2^l \quad \tau_2$ \\ 
\bottomrule
\end{tabular}
}
\caption{Hindi Sentence Structures Demonstrating Placement of \textit{Nahi} and Associated Pregroup Types. Note: superscripts `r' and `l' stand for right and left adjoints (inverses) of the pregroup types. The curved brackets and simple `@' are used for grouping multiple pregroup types that interact and evolve into a single output.}
\label{tab:nahi-structures}
\end{table}

\subsection{Incorporating the Negation Particle \textit{Nahi} in Hindi QNLP}

Sentiment analysis relies heavily on sentiment negation as one of its essential construction elements. The Hindi language uses the negation particle \textit{nahi} ("not"), which comes before the main verb of the clause~\cite{hindipod2021negation}. The finite verb receives \textit{nahi} immediately before it in basic sentence structures. 

For example, the sentence \textit{Ram nahi gaya} means ``Ram did not go,'' in contrast to the positive statement \textit{Ram gaya} which means ``Ram went.''

The pregroup representation allows the negation word to function as an adverbial modifier that maintains the original type reductions. The negation word \textit{nahi} operates by inserting itself into a specific verb phrase type and producing a verb phrase of the same type without modifying the required arguments of the verb. 

Negation can be incorporated by assigning \textit{nahi} a compound type of $xx.l$, where $x$ represents the type of the verb phrase it modifies so that \textit{nahi} $x$ maintains type $x$. The sentence \textit{Ram nahi gaya} receives the pregroup type sequence:
\begin{equation}
 n \;  ((n^r@s)(n^r@s)^l) \; n.r \; s,   
\end{equation}

which reduces to `$s$' in the same way as $n \; n.r \; s$ for affirmative cases.

The placement of \textit{nahi} appears before the main verb or participates in complex verb phrases that include auxiliaries. Table~\ref{tab:nahi-structures} presents various Hindi sentence structures with and without negation to demonstrate \textit{nahi} placement and the associated pregroup types.

\subsection{Manually Curated Hindi Dataset for Quantum Sentiment Classification}

We created a specific dataset for this study to evaluate sentiment classification using a QNLP framework for Hindi. The dataset contains 250 Hindi sentences that were manually created and annotated with positive, negative, or neutral sentiment labels.

The dataset was constructed and verified entirely by hand, keeping its size small to maintain high linguistic quality and structural accuracy. This method enabled strict grammatical accuracy and exact sentiment representation, validated by native Hindi speakers. The dataset functions as a proof-of-concept to demonstrate the QNLP pipeline, rather than striving for state-of-the-art performance on extensive corpora.

The vocabulary distribution across the sentiment classes was balanced to eliminate lexical bias, ensuring that classification depends on structural and semantic features rather than surface word frequency. The sentences contain between 5 to 7 words on average, making them suitable for building efficient quantum circuits.

\subsection{Sentence Composition}

The sentences were designed to reflect everyday statements that carry clear sentiment, including emotionally expressive and neutral factual expressions:

\begin{itemize}
    \item \textbf{Positive Sentences:} Expressions of praise, happiness, etc.\\
    Example: \textit{Mujhe tumari saari baatein pasand hai} (``I like everything about you.'')
    
    \item \textbf{Negative Sentences:} Expressions of complaints, sadness, etc.\\
    Example: \textit{Ve farsh nahi saaf kar rahe} (``They are not cleaning the floor.'')
    
    \item \textbf{Neutral Sentences:} Objective or emotionless factual statements.\\
    Example: \textit{Gautami ne khana banaya} (``Gautami cooked food.'')
\end{itemize}

These sentences ensure that the model learns to distinguish between emotional and non-emotional content based on structure and semantics, not just keywords.

The pregroup type assignments accompany each sentence through the Hindi grammar lexicon described in the methodology section. The dataset includes syntactic diversity across common constructions, such as copula forms, transitive verb phrases, and negated predicates.

Given its size, the dataset was not intended to train a high-capacity model but to serve as a proof-of-concept corpus. Every sentence was manually checked by native Hindi speakers to confirm grammatical accuracy and correct sentiment expression. Vocabulary distribution remained equal among sentiment classes to avoid lexical bias and ensure that classification relied on compositional structure. 

\begin{figure*}[htbp]
  \centering
  \resizebox{\textwidth}{!}{\includegraphics{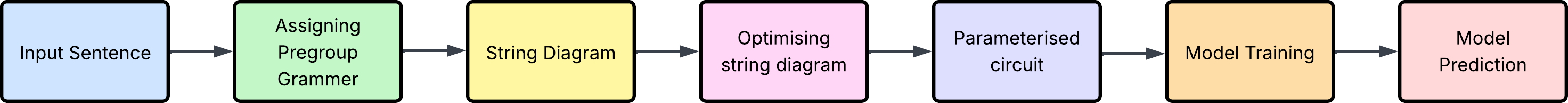}}
  \caption{Lambeq Pipeline}
  \label{pipeline}
\end{figure*}

\section{Methodology}

\subsection{Dataset}
The dataset contains a collection of sentences with varying sentiments, including both positive, negative, and neutral examples. The dataset was divided into training, validation, and test sets in a ratio of 70:10:20. The dataset comprised entirely single, grammatically complete sentences. A restricted vocabulary was employed to decrease the number of unfamiliar words within the test sets without total elimination. This approach ensures some degree of generalisability over the test words while maintaining a wealth of common vocabulary.

\subsection{QNLP Pipeline}
The QNLP pipeline for the Hindi sentiment classification is shown in Figure 1. The pipeline starts with a Hindi sentence as input, parses it into a pregroup grammar representation, converts it into a string diagram, translates the diagram into a parameterized quantum circuit, and finally integrates the circuit into a hybrid quantum-classical model. In summary, our pipeline comprises four main stages (illustrated in  Figure~\ref{pipeline}):

\textbf{1. Sentence to String Diagram: }The process begins by applying the Hindi pregroup grammar to each sentence to assign atomic types to every  word. The grammar assigns the noun type denoted as `n' to all noun words while transitive verbs  receive compound types that transform two noun inputs into a sentence output `s'. The DisCoCat parser  from Lambeq transforms grammar reductions into string diagrams which depict word types as wires and implement  noun-adjoint type contractions through cup-shaped connections. The cups exist as entangled wire pairs that connect the words in the  diagram~\cite{srivastava2023enabling}.

  \textbf{2. Diagram Simplification:} The Lambeq tool simplifies diagrams through its application of rewrite rules to optimize string diagrams. The optimization process removes superfluous cups and decreases the wire complexity (such as removing duplicate noun wires) while preserving the original sentence meaning. The reduction of unnecessary cups helps decrease the quantum state dimensionality, which leads to fewer qubits and gates required in the circuit.

\textbf{3. String Diagram to Quantum Circuit:} The simplified diagram is then converted into a parameterized quantum circuit  (PQC) using a chosen ansatz. We employ Lambeq’s built-in Instantaneous Quantum Polynomial (IQP) ansatz for this conversion. In this step, each wire in the diagram (of types like  `n' or `s') is mapped to one or more qubits, and each diagrammatic operation  (ex. each word or cup) is mapped to a sequence of quantum gates. For example, each word state is prepared by a parameterized gate sequence on its qubit(s), and each cup  (which connects a noun to a verb’s adjoint type) is implemented by an entangling gate  (such as a controlled-Z rotation) between the corresponding qubits. The result is a variational quantum circuit whose adjustable gate parameters \(\theta_i\) correspond to the weights of the word meanings in the sentence.\\
  \vspace{-0.5mm}
 \textbf{4. Hybrid Model Training:} The hybrid model training process uses generated circuits for each sentence as part of a sentiment classification system. The quantum circuit receives trainable parameters for each sentence before its execution (simulation) to generate output results such as measurement probabilities or expectation values. The quantum output data flows into a classical post-processing layer (Section 3.3) to generate the final prediction. The model receives its end-to-end training through the evaluation of predicted sentiment against actual sentiment labels and the calculation of loss. The circuit parameters receive updates through gradient descent using a classical optimizer from Lambeq's PyTorch backend with the Adam optimizer until the model achieves convergence~\cite{kartsaklis2021lambeqefficienthighlevelpython}.
\subsection*{Example: Pregroup Parsing and Circuit}

We demonstrate the approach by analyzing the Hindi sentence \textbf{``Gautami faal khati hai''}, which means \textit{``Gautami eats fruit.''} The sentence contains a subject (Gautami) and an object (fruit) together with a verb phrase (``eats'').

We begin by assigning each word its corresponding type from the Hindi pregroup grammar as shown below:

\vspace{0.5cm}

\noindent
\textbf{Sentence:} ``Gautami faal khati hai''

\vspace{0.5cm}

\noindent
\textbf{Types:}
\begin{align*}
\text{Gautami} &\rightarrow \pi_{13} \\
\text{faal} &\rightarrow o \\
\text{khati} &\rightarrow o^{r} \cdot \pi_{13}^{r} \cdot s \cdot \tau_{2}^{l} \\
\text{hai} &\rightarrow \tau_{2}
\end{align*}

\begin{figure}[htbp]
  \centering
  \resizebox{0.55\textwidth}{!}{\includegraphics{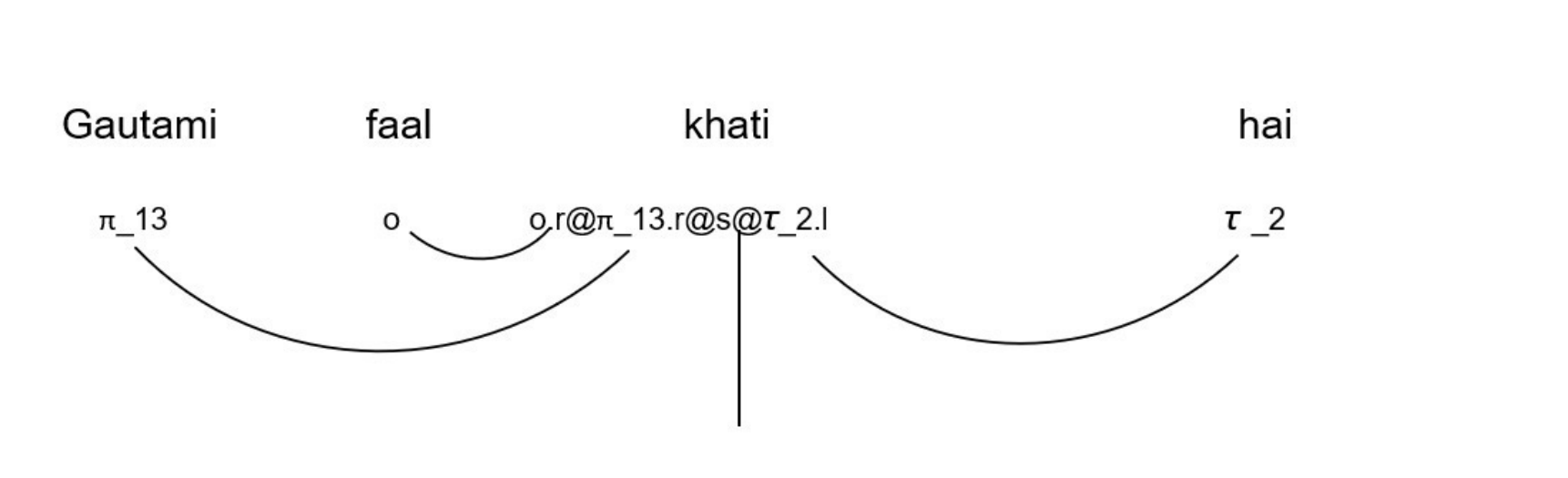}}
  \caption{Pregroup. Note: Labels like $\pi_{13}$, $\tau_{21}$ are generated from the Lambeq code, and appear here as $\pi\_13$ and $\tau\_21$, same is with other labels.}
  \label{pregroup_example_sentence}
\end{figure}

\subsubsection*{Step 1: Creating the Sentence Diagram}

\begin{figure}[htbp]
  \centering
  \resizebox{0.45\textwidth}{!}{\includegraphics{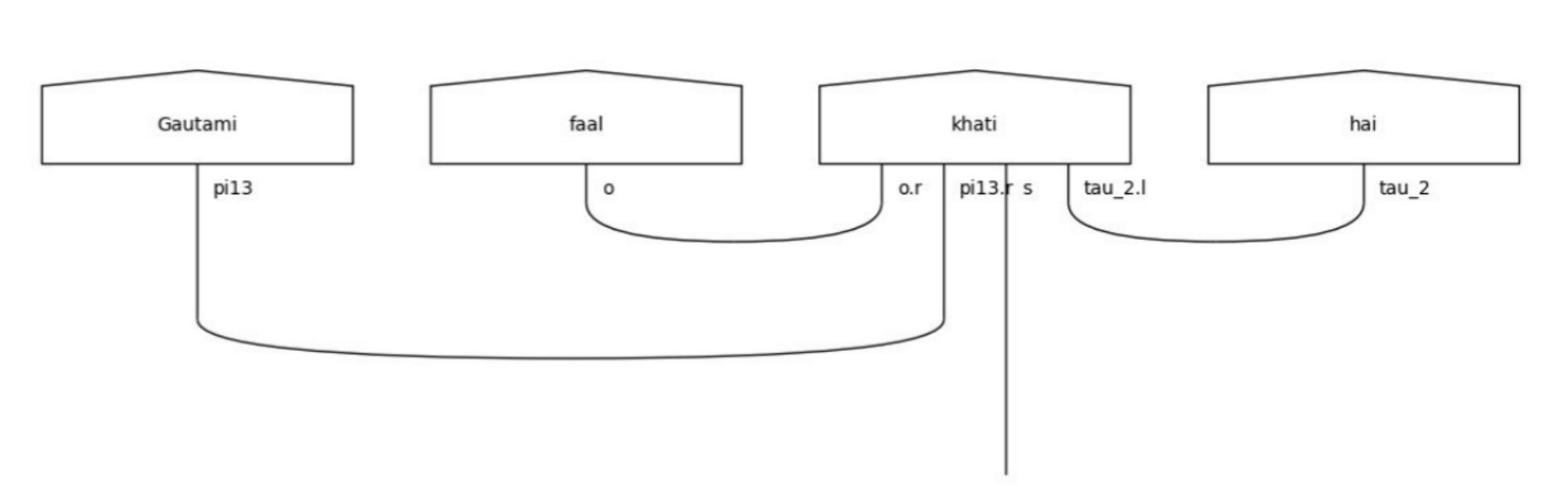}}
  \caption{String Diagram. Note: Labels like $\pi_{13}$, $\tau_{21}$ are generated from the Lambeq code, and appear here as pi13 and tau\_21, same is with other labels.}
  \label{string}
\end{figure}

Using pregroup notation and word connections, we apply the pregroup grammar framework to describe Hindi. Sentence diagrams are created for the given sentences as shown in sentence diagram of the Figure~\ref{pregroup_example_sentence}. As stated in Section 4.1, the sentence yields the corresponding string diagram shown in the Figure~\ref{string}.

\subsubsection*{Step 2: Building the Parameterized Quantum Circuit}

The simplified diagram is translated into a parameterized quantum circuit through Lambeq's circuit compiler. The atomic wires (such as $\pi_{13}$, $o$, and $\tau_{2}$) function as qubits in the system, and the word boxes transform into parameterized unitary blocks that contain trainable quantum gate operations.

\vspace{0.4cm} % small vertical space before bullets

The process involves:

\begin{itemize}
    \item The embedding of \textit{Gautami} and \textit{faal} occurs through $R_z(\theta)$ layers, which perform single-qubit state preparation.
    \item The \textit{khati} operation functions as a multi-qubit block that creates entanglement between the qubits representing $\pi_{13}$, $o$, and $\tau_{2}$.
    \item The operator \textit{hai} functions as a one-qubit operator that targets $\tau_{2}$.
    \item The diagram's cups are implemented through entangling gates (typically CZ gates) that connect the corresponding qubits according to grammatical rules.
\end{itemize}

The resulting quantum circuit maintains the compositional meaning of the sentence while incorporating trainable parameters $\{\theta_i\}$, which are optimized during the classification process.

\begin{figure*}[htbp]
  \centering
  \resizebox{0.9\textwidth}{!}{\includegraphics{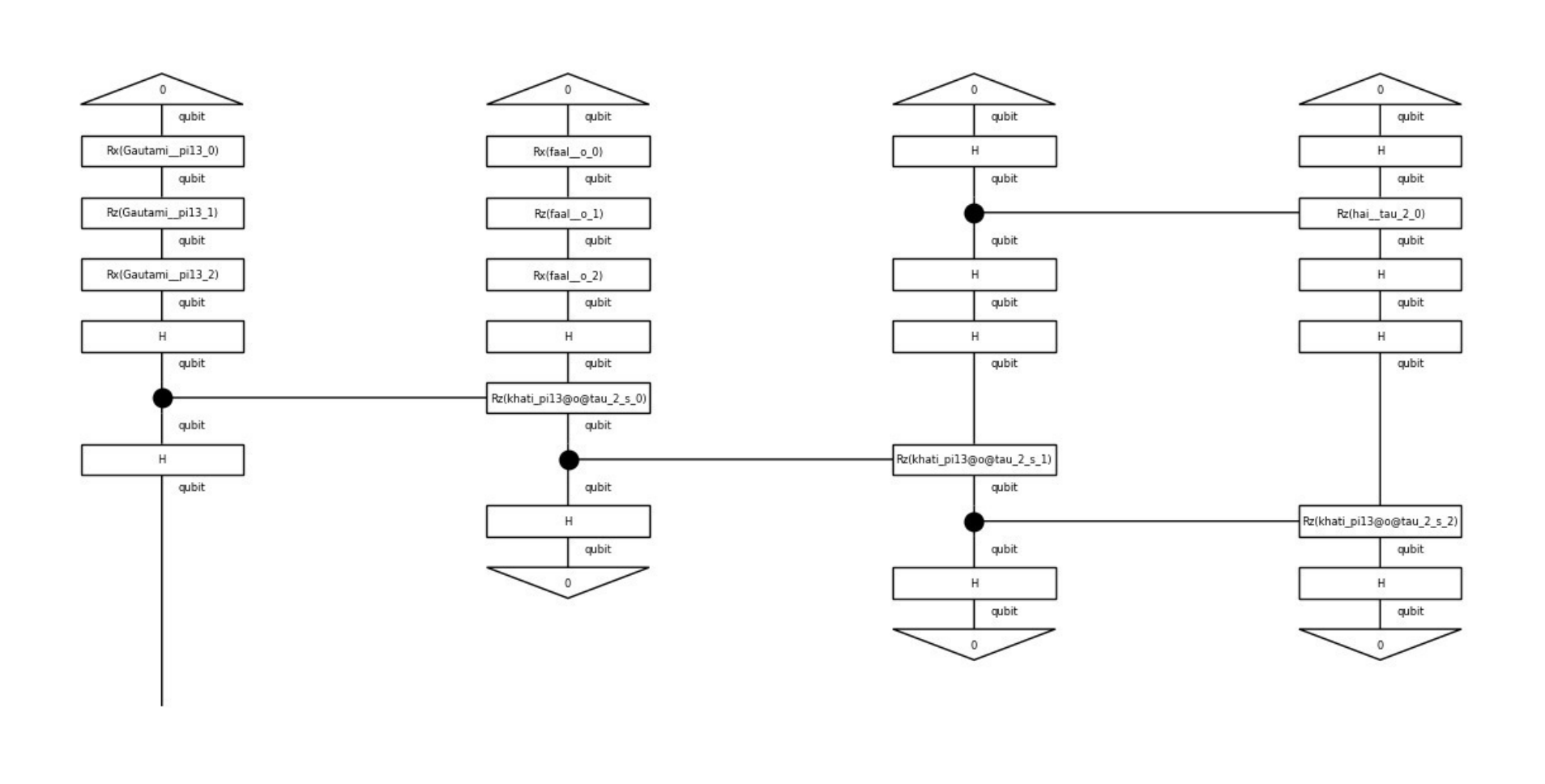}}
  \caption{Quantum Circuit. Note: Labels like $\pi_{13}$, $\tau_{21}$ are generated from the Lambeq code, and appear here as pi\_13 and tau\_21, same is with other labels.}
  \label{circuit}
\end{figure*}

\subsubsection*{Step 3 IQP Ansatz}

The IQP (Instantaneous Quantum Polynomial-time) ansatz built into Lambeq is used to construct the circuits. This ansatz consists of:
\begin{itemize}
    \item Hadamard gates applied to all qubits
    \item Followed by parameterized $R_z$ rotations
    \item Entangling CZ gates between qubits as dictated by the grammatical cups
\end{itemize}
Each word contributes a local subcircuit, and the entire circuit is variational, with parameters trained to encode the sentiment meaning of the sentence. See Figure~\ref{circuit} for the circuit corresponding to the example sentence
\section{Results and Discussion}

\begin{figure*}[htbp]
  \centering
  \resizebox{\textwidth}{!}{\includegraphics{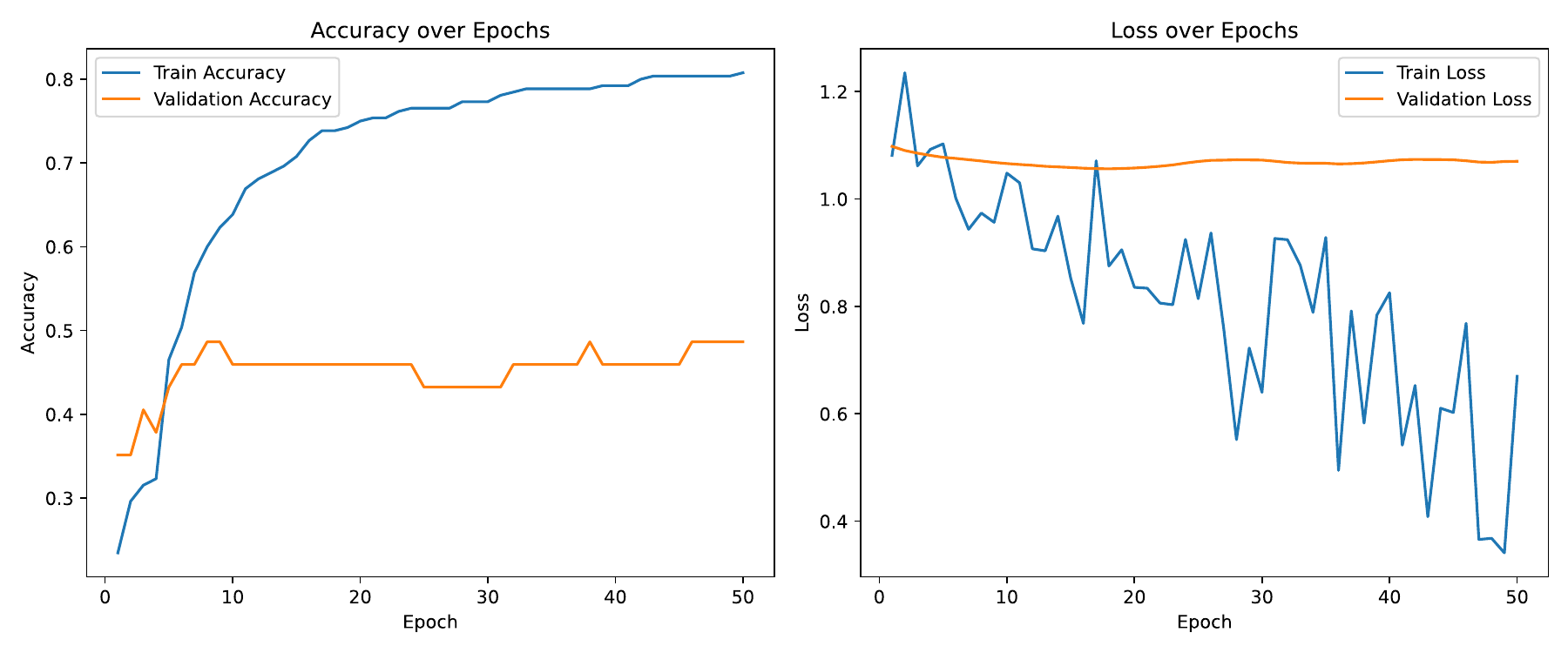}}
  \caption{Ternary Class Classification. Left: Accuracy Vs Epoch, Right: Loss Vs Epoch.}
  \label{ternary}
\end{figure*}

\begin{figure*}[htbp]
  \centering
  \resizebox{\textwidth}{!}{\includegraphics{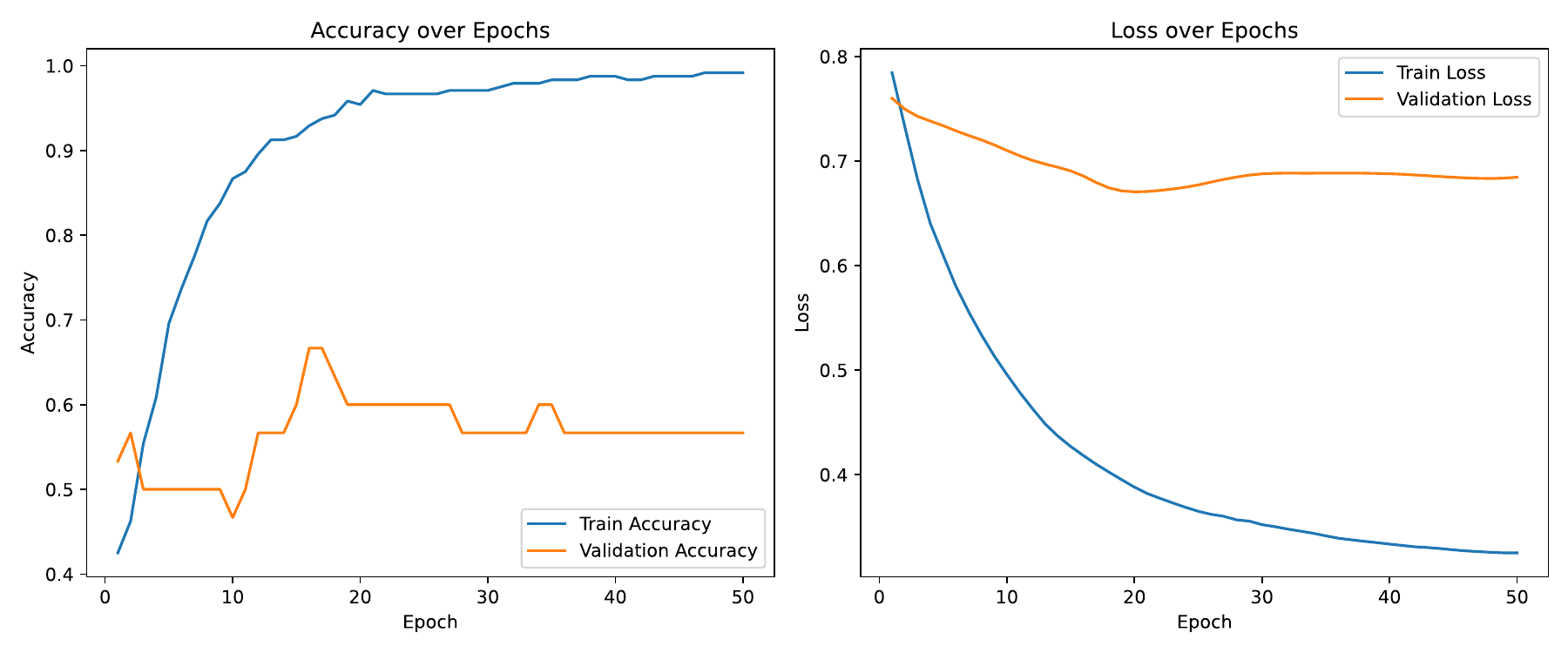}}
  \caption{Binary Class Classification. Left: Accuracy Vs Epoch, Right: Loss Vs Epoch.}
  \label{binary}
\end{figure*}

The performance of the proposed hybrid quantum-classical model was evaluated across two sentiment classification tasks: a \textbf{Ternary-class task} (positive, negative, neutral) and a \textbf{Binary-class task} (positive, negative). Additionally, experiments were conducted using two data split configurations: \textbf{70:10:20} and \textbf{80:10:10} (train: validation: test). All models were trained and tested using simulated quantum circuits.

\begin{table}[h]
\centering
\begin{tabular}{|l|c|c|}
\hline
\textbf{Task Type} & \textbf{70:10:20 Split} & \textbf{80:10:10 Split} \\
\hline
Ternary-Class Classification & 55.0 & 61.0 \\
Binary-Class Classification & 68.0 & 73.0 \\
\hline
\end{tabular}
\caption{Sentiment Classification Accuracy (\%)}
\label{tab:accuracy and loss}
\end{table}

The dataset used in this study contained a highly diverse vocabulary, with many words appearing only once or a few times. While this variety proves the working of the model, it also presents challenges for learning and generalization. The model was particularly affected in the Ternary-class classification setting, where low-frequency words offered limited learning opportunities for embedding representations. 
As shown in Table~\ref{tab:accuracy and loss}, performance in the Binary-class classification task was consistently stronger than in the Ternary-class version, highlighting the difficulty in accurately detecting neutral sentiment, which often lacks distinctive emotional or grammatical features. Additionally, the comparison of dataset splits reveals that the 80:10:10 configuration yielded higher accuracy than the 70:10:20 split for both tasks. These results are substantiated in the Accuracy and Loss Vs Epoch plots shown in the Figures~\ref{ternary} \& \ref{binary}. This suggests that a larger training set provides greater stability during quantum parameter optimization, indicating that hybrid quantum models are notably sensitive to training data volume and may underfit with limited samples. 

\section{Conclusion}
 The research presented a grammar-aware QNLP pipeline for Hindi sentiment classification via the application of compositional quantum models to complex morphological languages. The pregroup grammar framework enabled us to construct sentence diagrams that integrated the Hindi negation particle {\em "nahi"} directly into the syntactic structure. The Lambeq toolkit converts these diagrams into parameterized quantum circuits. The hybrid quantum-classical neural network  (HQNN) demonstrated promising accuracy in both binary (positive/negative) and ternary  (positive/negative/neutral) sentiment classification tasks when evaluated on a manually annotated corpus of 250 Hindi sentences. The compositional QNLP method successfully detects Hindi semantic features, including intricate morphological patterns and negation-induced semantic transformations.

This research presents the initial implementation of complete compositional QNLP methods for Hindi that possess complex morphology and adaptable word arrangement. The proposed model successfully handles the  {\em "nahi"} negation particle, which represents a significant achievement because negation has traditionally been difficult to analyze in sentiment analysis. The proposed approach has the advantage of grammar-informed quantum models, which structurally represent negation in quantum circuits to properly detect changes in sentiment polarity. The proposed Hindi QNLP sentiment classifier demonstrates that compositional quantum techniques can be applied to complex linguistic structures beyond English language boundaries. The research expands QNLP applications and demonstrates how sentiment analysis models can become more transparent through linguistic foundations.

\section{Future Scope}
 We have several plans to extend this research. First, we increase the size and diversity of the Hindi dataset to improve the model’s generalizability. A larger corpus with more varied vocabulary and syntactic constructions, including interrogatives, imperatives, and complex compound sentences, will allow the proposed HQNN to learn a broader range of linguistic patterns and better handle edge cases. The inclusion of more diverse sentence types and grammatical structures (e.g., different word orders or embedded clauses) will further strengthen the compositional framework and ensure its robustness in capturing Hindi’s rich syntax.

Second, we plan to apply the proposed QNLP pipeline to other Indian languages, such as Telugu, Kannada, and Marathi. These languages belong to different linguistic families and feature distinct grammatical features. By adapting the pregroup grammar diagrams and quantum circuit ansätze to these languages, we can evaluate the generality of our approach across diverse morphosyntactic settings. The successful implementation of multiple languages demonstrates that compositional QNLP techniques are broadly applicable and not limited to a single linguistic context.

Finally, the ultimate goal is to deploy and test the proposed model on actual quantum hardware. To date, all experiments have been conducted on simulated or hybrid quantum platforms; running the sentiment classification circuits on real quantum processors is a crucial step toward validating the practicality of the proposed approach. As quantum computing technology matures, we aim to assess the performance of the proposed HQNN on noisy intermediate-scale quantum (NISQ) devices and identify any quantum-specific advantages or challenges. This transition from simulation to physical quantum hardware provides deeper insights into the feasibility of QNLP for real-world applications and helps guide further refinement of the model.
\bibliographystyle{splncs04}
\bibliography{cite}




% LNCS Homepage, \url{http://www.springer.com/lncs}, last accessed 2023/10/25
% \end{thebibliography}
\end{document}